\documentclass{mn2e}
\usepackage{graphicx}

\title[Orbit, masses and distance to $\beta$ Centauri]{Orbital parameters,
masses and distance to $\beta$ Centauri determined with the Sydney
University Stellar Interferometer and high resolution
spectroscopy}

\author[J. Davis et al.]{J. Davis,$^1$\thanks{E-mail: davis@physics.usyd.edu.au}
 A. Mendez,$^1$ E.B. Seneta,$^1$\thanks{Present address: Astrophysics Group,
Cavendish Laboratory, Madingley Road, Cambridge CB3 0HE, UK} W.J.
Tango,$^1$ A.J. Booth,$^1$\thanks{Present address: Jet Propulsion
Laboratory, California Institute of Technology, Pasadena, CA
91109, USA} \newauthor J.W. O'Byrne,$^1$  E.D.
Thorvaldson,$^1$\thanks{CSIRO Division of Telecommunications \&
Industrial Physics, West Lindfield, NSW 2070, Australia} M.
Ausseloos,$^2$ C. Aerts,$^{2,\,3}$ \newauthor K. Uytterhoeven$^2$\\
$^1$ School of Physics, University of Sydney, NSW 2006, Australia\\
$^2$ Institute of Astronomy , Catholic University of Leuven,
Celestijnenlaan 200 B, 3001 Leuven, Belgium\\
$^3$ Department of Astrophysics, University of Nijmegen, PO Box
9010, 6500 GL Nijmegen, The Netherlands }

\input{psfig}

\begin{document}

\maketitle

\begin{abstract}
The bright southern binary star $\beta$ Centauri (HR5267) has been
observed with the Sydney University Stellar Interferometer (SUSI)
and spectroscopically with the ESO CAT and Swiss Euler telescopes
at La Silla. The interferometric observations have confirmed the
binary nature of the primary component and have enabled the
determination of the orbital parameters of the system.  At the
observing wavelength of 442\,nm the two components of the primary
system have a magnitude difference of 0.15$\pm$0.02.  The
combination of interferometric and spectroscopic data gives the
following results: orbital period $357.00\pm0.07$\,days,
semi-major axis $25.30\pm0.19$\,mas, inclination
67.4$\pm$0.3\,degrees, eccentricity 0.821$\pm$0.003, distance
$102.3\pm1.7$\,pc, primary and secondary masses $M_{1} = M_{2} =
9.1\pm0.3$\,M$_{\odot}$ and absolute visual magnitudes of the
primary and secondary $M_{1V} = -3.85\pm0.05$ and $M_{2V} =
-3.70\pm0.05$.  The high accuracy of the results offers a fruitful
starting point for future asteroseismic modelling of the pulsating
binary components.
\end{abstract}


\section{Introduction}

The bright southern star $\beta$ Centauri (HR 5267) is classed in
the {\em Bright Star Catalogue} \cite{82bsc} as spectral type
B1\,III, and as a visual double of separation 1\farcs3 with a
magnitude difference of 3.2.  In 1963, during the early
commissioning observations with the Narrabri Stellar Intensity
Interferometer (NSII) \cite{67arhbetal}, it was found that the
instrument's response to $\beta$ Cen was approximately half that
expected from a single star. From this observation it was deduced
that the primary of the visual double was in fact a binary system
with two roughly equal components \cite{74hbdanda}.  It was not
possible to determine additional information about $\beta$ Cen
with the NSII, but support for the binary nature of the primary
came from the spectroscopic observations of Breger
\shortcite{67breger}, Shobbrook and Robertson \shortcite{68sandr},
and Lomb \shortcite{75nl}.  They found that the primary probably
contained a $\beta$ Cephei variable with period of either 0.135\,d
or 0.157d, and that residual velocities for the lines are
consistent with a binary which Shobbrook and Robertson suggested
had a period of 352\,d. The spectroscopic observations led to
$\beta$ Cen being classified as a single-lined spectroscopic
binary but Robertson et al. \shortcite{99dlsb} have used aperture
masking interferometry and high resolution spectroscopy to confirm
that the primary is indeed a binary, to resolve the angular
separation of the two components, and to establish it as a
double-lined spectroscopic binary.  Ausseloos et al.
\shortcite{aus02} have made a detailed spectroscopic study of
$\beta$ Cen and shown that it is an eccentric binary with two
$\beta$ Cep-type components and determined the orbital parameters
of the system.

In this paper we report observations of the $\beta$ Cen system
made with the Sydney University Stellar Interferometer (SUSI)
\cite{99susi1} which not only confirm the detection of the two
components of the primary reported by Robertson et al. but also
enable the orbital parameters of the system to be determined.  In
this paper `$\beta$ Cen' will be taken to identify the
spectroscopic binary alone unless the fainter visual companion is
specifically mentioned.  We also report an update of the
spectroscopic study of $\beta$ Cen by Ausseloos et al.
\shortcite{aus02} and combine the interferometric and
spectroscopic results to determine the distance to the system and
the masses and absolute magnitudes of the components.

\section{Interferometric Observations}\label{sec:obs}

SUSI is a long baseline optical stellar interferometer with a
North-South baseline array.  It measures the square of the fringe
visibility which we term the `correlation' $C$ \cite{99susi2}.
The correlation for a binary star is given by

\begin{equation}
C = \frac{1}{(1 + \beta)^{2}}\left[\Gamma_{1}^{2} + \beta^{2}\Gamma_{2}^{2} + 2\beta|\Gamma_{1}||\Gamma_{2}|\cos(\alpha)\right]
\label{eq:cor}
\end{equation}

\noindent where $\beta$ is the brightness ratio of the two components (defined
so that $\beta\leq 1$) and

\begin{eqnarray}
\Gamma_{1} & = & \frac{2J_{1}(\pi\theta_{UD1}b/\lambda_{0})}{\pi\theta_{UD1}b/\lambda_{0}} \label{eq:theta1} \\
\Gamma_{2} & = & \frac{2J_{1}(\pi\theta_{UD2}b/\lambda_{0})}{\pi\theta_{UD2}b/\lambda_{0}} \label{eq:theta2} \\
\alpha    & = & \frac{2\pi b\rho\cos{\psi}}{\lambda_{0}}
\label{eq:costerm}
\end{eqnarray}

The symbols in equations~(\ref{eq:theta1}) and (\ref{eq:theta2}) are defined
as follows: $b$ is the length of the interferometer baseline projected onto
the sky, $\lambda_{0}$ is the observing wavelength, $\theta_{UD1}$ and $\theta_{UD2}$
are the equivalent uniform-disk angular diameters of the primary and secondary
components, and $J_{1}(x)$ is a Bessel function.  In equation~(\ref{eq:costerm})
$\rho$ is the angular separation of the two components and $\psi$ is the angle
between the line joining the two stars and the projection of the baseline onto
the sky.

The rotation of the Earth causes both $b$ and $\psi$ to vary with time.  $\psi$
will also vary due to orbital motion of the binary system.  The result of
these variations is to cause the cosine term in equation~(\ref{eq:cor}), and
hence the observed correlation, to vary with hour angle.  The timescale of the
variation in correlation depends on the baseline employed and, for each
observing night, the baseline was chosen to give at least one, and preferably
two minima in the observed correlation.

In order to obtain accurate measurements of the correlation it it
essential to ensure that any difference in optical path length
$\Delta$OPL between the two arms of the interferometer is small
compared with the coherence length of the interferometer. In SUSI
this is done by means of an optical path length compensator (OPLC)
\cite{99susi1}. Ideally an automatic fringe tracker would be used
to control the OPLC and maintain path equality but, at the time of
the observations reported here, this was not available and the
following technique was employed.  The correlation was measured
for several values of the optical path difference between the two
arms of the interferometer bracketing the matched path position.
The correlation at $\Delta$OPL = 0 was found by fitting a `delay'
curve to the data \cite{99susi2}.  A series of delay curves was
obtained for each night to give a time series of correlation
values for $\Delta$OPL = 0.

The usual mode of operation for interferometric observations is to
interleave observations of the programme star or binary system
with observations of calibrators, stars that are either unresolved
or of known angular diameter.  The limiting magnitude of SUSI at
the time of the observations was B $\sim$ 2.5 and there were no
suitable calibrators in the vicinity of $\beta$ Cen.  The observed
correlation from $\beta$~Cen was generally changing rapidly due to
the cosine term in equation~(\ref{eq:cor}) and it was desirable to
monitor it as continuously as possible.  In the absence of
suitable calibrators and the lack of a fringe-tracking system it
was decided to omit observations of calibration sources and,
instead of endeavouring to measure the angular sizes of the
components as well, to concentrate on the determination of the
orbital elements of the system.  This omission adds some
complication to the analysis but has minimal implications for the
determination of the separations and position angles of the
component stars.  With these considerations in mind, the
observations were made at a wavelength of 442\,nm and with
baseline lengths of 5\,m, 10\,m, 15\,m, 20\,m or 40\,m.  At these
baselines the two component stars of the system are only partially
resolved and the correlation for either of the components alone is
estimated to be $\geq0.95$ for all the baselines except 40\,m
where the correlation is estimated to be $>$0.8 (see
Section~\ref{sec:analysis}).

A total of 43 nights of observations of $\beta$~Cen were made with
SUSI during the period April 1997 to April 2002.  The dates of the
observations and the baselines used are listed in
Table~\ref{tab:vectors}.

\begin{table*}
\begin{minipage}{110mm}
  \caption{Measured angular separations ($\rho$) and
   position angles ($\theta$) for $\beta$ Cen.  $\Delta\rho$ and $\Delta\theta$ are the
   differences between the measured values and the corresponding $\rho$ and $\theta$
   values for the fitted orbit.  The weights are those assigned for the orbital fitting
   program (see Section~\ref{sec:orbit}). Further details are given in the text.}
  \label{tab:vectors}
  \begin{tabular}{rcrrrrrc}
\hline
\multicolumn{1}{c}{Date} & Mean & \multicolumn{1}{c}{B} & \multicolumn{1}{c}{$\rho$} & \multicolumn{1}{c}{$\Delta\rho$}& \multicolumn{1}{c}{$\theta$} & \multicolumn{1}{c}{$\Delta\theta$} & Weight\\
\multicolumn{1}{c}{(UT)} & MJD & \multicolumn{1}{c}{(m)} & \multicolumn{1}{c}{(mas)} & \multicolumn{1}{c}{(mas)} & \multicolumn{1}{c}{(deg)} & \multicolumn{1}{c}{(deg)} & \\
\hline
27/4/97 & 50565.66 & 10 & 21.69 $\pm$ 0.20 & -0.05 & 118.07  $\pm$ 0.16 &  0.14 & 3 \\
8/5/97  & 50576.56 & 5  & 24.83 $\pm$ 0.39 &  0.48 & 120.54  $\pm$ 0.52 & -0.24 & 3 \\
9/5/97  & 50577.59 & 10 & 24.03 $\pm$ 0.19 & -0.53 & 121.52  $\pm$ 0.34 &  0.49 & 4 \\
10/5/97 & 50578.55 & 5  & 25.94 $\pm$ 0.39 &  1.19 & 119.73  $\pm$ 0.42 & -1.52 & 4 \\
29/5/98 & 50962.57 & 20 & 28.07 $\pm$ 0.05 & -0.37 & 126.65  $\pm$ 0.05 &  0.14 & 5 \\
30/5/98 & 50963.53 & 10 & 28.00 $\pm$ 0.06 & -0.52 & 127.45  $\pm$ 0.11 &  0.77 & 2 \\
8/6/98  & 50972.50 & 5  & 28.94 $\pm$ 0.35 & -0.24 & 128.11  $\pm$ 0.23 & -0.07 & 4 \\
9/6/98  & 50973.50 & 10 & 28.82 $\pm$ 0.06 & -0.42 & 128.53  $\pm$ 0.07 &  0.19 & 5 \\
23/7/98 & 51017.45 & 10 & 29.68 $\pm$ 0.25 & -0.14 & 135.33  $\pm$ 0.27 &  0.09 & 4 \\
20/3/99 & 51257.64 & 20 & 12.63 $\pm$ 0.06 & -0.15 & 107.54  $\pm$ 0.15 & -0.04 & 5 \\
25/3/99 & 51262.63 & 20 & 15.79 $\pm$ 0.11 &  0.30 & 109.73  $\pm$ 0.25 & -1.35 & 4 \\
26/3/99 & 51263.62 & 20 & 15.81 $\pm$ 0.05 & -0.15 & 111.52  $\pm$ 0.10 & -0.12 & 5 \\
27/3/99 & 51264.52 & 20 & 16.37 $\pm$ 0.25 &  0.00 & 111.59  $\pm$ 0.61 & -0.52 & 3 \\
29/3/99 & 51266.61 & 20 & 17.16 $\pm$ 0.06 & -0.12 & 112.88  $\pm$ 0.10 & -0.26 & 5 \\
6/4/99  & 51274.67 & 20 & 20.06 $\pm$ 0.04 & -0.19 & 116.27  $\pm$ 0.08 & -0.07 & 5 \\
10/4/99 & 51278.75 & 15 & 21.08 $\pm$ 0.10 & -0.41 & 117.45  $\pm$ 0.05 & -0.20 & 4 \\
11/4/99 & 51279.62 & 15 & 21.57 $\pm$ 0.04 & -0.17 & 117.84  $\pm$ 0.08 & -0.07 & 5 \\
12/4/99 & 51280.60 & 15 & 21.75 $\pm$ 0.03 & -0.25 & 118.14  $\pm$ 0.05 & -0.05 & 5 \\
18/4/99 & 51286.64 & 15 & 23.30 $\pm$ 0.05 & -0.21 & 119.84  $\pm$ 0.05 &  0.02 & 5 \\
19/4/99 & 51287.58 & 10 & 23.18 $\pm$ 0.12 & -0.53 & 120.39  $\pm$ 0.14 &  0.34 & 4 \\
22/4/99 & 51290.59 & 10 & 24.38 $\pm$ 0.13 &  0.02 & 120.66  $\pm$ 0.14 & -0.12 & 4 \\
23/4/99 & 51291.59 & 5  & 25.09 $\pm$ 0.46 &  0.53 & 120.47  $\pm$ 0.50 & -0.54 & 4 \\
28/4/99 & 51296.60 & 10 & 25.52 $\pm$ 0.20 &  0.03 & 121.98  $\pm$ 0.24 & -0.15 & 4 \\
29/4/99 & 51297.60 & 5  & 26.57 $\pm$ 0.37 &  0.90 & 121.51  $\pm$ 0.33 & -0.84 & 3 \\
1/5/99  & 51299.62 & 5  & 26.59 $\pm$ 0.37 &  0.59 & 121.86  $\pm$ 0.32 & -0.91 & 3 \\
13/5/99 & 51311.55 & 10 & 27.84 $\pm$ 0.14 &  0.18 & 125.23  $\pm$ 0.20 &  0.15 & 3 \\
14/5/99 & 51312.55 & 10 & 28.17 $\pm$ 0.15 &  0.43 & 124.94  $\pm$ 0.27 & -0.32 & 3 \\
18/5/99 & 51316.56 & 10 & 28.29 $\pm$ 0.14 &  0.13 & 125.80  $\pm$ 0.15 & -0.18 & 4 \\
28/5/99 & 51326.55 & 10 & 28.83 $\pm$ 0.10 & -0.16 & 128.01  $\pm$ 0.13 &  0.32 & 4 \\
1/6/99  & 51330.52 & 10 & 29.08 $\pm$ 0.06 & -0.16 & 129.27  $\pm$ 0.13 &  0.93 & 5 \\
16/6/99 & 51345.47 & 10 & 29.99 $\pm$ 0.12 &  0.13 & 130.65  $\pm$ 0.16 & -0.08 & 4 \\
21/6/99 & 51350.53 & 10 & 29.48 $\pm$ 0.27 & -0.48 & 132.64  $\pm$ 0.28 &  1.12 & 3 \\
28/6/99 & 51357.48 & 10 & 30.33 $\pm$ 0.18 &  0.31 & 133.21  $\pm$ 0.27 &  0.62 & 3 \\
28/7/99 & 51387.46 & 10 & 29.36 $\pm$ 0.33 &  0.00 & 137.39  $\pm$ 0.40 &  0.08 & 3 \\
29/7/99 & 51388.43 & 10 & 29.14 $\pm$ 0.14 & -0.17 & 137.44  $\pm$ 0.15 & -0.03 & 4 \\
10/8/99 & 51400.43 & 10 & 28.14 $\pm$ 0.30 & -0.52 & 139.87  $\pm$ 0.32 &  0.41 & 3 \\
28/1/00 & 51571.71 & 40 &  8.81 $\pm$ 0.05 & -0.11 & 235.83  $\pm$ 0.12 & -0.89 & 3 \\
2/2/00  & 51576.68 & 40 &  8.60 $\pm$ 0.09 & -0.12 & 245.10  $\pm$ 0.12 & -0.55 & 3 \\
6/2/00  & 51580.69 & 40 &  8.26 $\pm$ 0.25 & -0.29 & 252.60  $\pm$ 0.25 & -0.51 & 2 \\
16/2/00 & 51590.71 & 40 &  7.20 $\pm$ 0.09 & -0.34 & 273.83  $\pm$ 0.20 & -0.37 & 3 \\
18/2/00 & 51592.69 & 40 &  6.82 $\pm$ 0.04 & -0.20 & 278.98  $\pm$ 0.16 & -0.40 & 5 \\
2/3/00  & 51605.63 & 40 &  5.67 $\pm$ 0.08 & -0.16 &  91.51  $\pm$ 0.31 & -0.16 & 5 \\
14/4/02 & 52378.66 & 10 & 26.57 $\pm$ 0.15 & -0.58 & 124.53  $\pm$ 0.15 &  0.19 & 4 \\
 \hline
\end{tabular}
\end{minipage}
\end{table*}

\section{Analysis of the Interferometric Observations} \label{sec:analysis}
The measured values of correlation are not the true values of
correlation due to instrumental and seeing losses. Temporal seeing
effects are corrected on-line by the SUSI data acquisition system
\cite{96dandt,99susi2} but spatial seeing effects and instrumental
losses could not be corrected in the normal way in the absence of
calibration source observations. However, since the aim is to
determine the separation and position angle of the binary, the
absolute values of the correlation are not important.  Even for
the determination of the brightness ratio $\beta$, it is only the
ratio of maximum to minimum correlation that must be determined
and this is discussed in Section~\ref{sec:beta}.

A least squares fit of equation~(\ref{eq:cor}) with $\Gamma_{1} =
\Gamma_{2} = \gamma$ was made to the data for each night with
$\gamma$, $\rho$ and $\theta$ as free parameters where $\theta$ is
the position angle of the binary system (measured east from
north).  The justification for the assumption of $\Gamma_{1} =
\Gamma_{2}$ or, in other words $\theta_{UD1} = \theta_{UD2}$, was
initially based on the finding of Hanbury Brown, Davis \& Allen
\shortcite{74hbdanda} that the two component stars of $\beta$~Cen
are of comparable brightness.  Following the determination of the
brightness ratio $\beta$ (Section~\ref{sec:beta}) the angular
diameters of the primary and secondary were estimated to be
0.77\,mas and 0.72\,mas respectively based on the angular
diameters of early-type stars by Hanbury Brown, Davis \& Allen
\shortcite{74hbdanda}.  Based on these estimates, the correlation
for either of the components is $\geq$0.95 for all the baselines
employed except for 40\,m where the correlation is $>$0.8.  For
some 67\% of the observations it follows that $\Gamma_{1}$ and
$\Gamma_{2}$ differ by less than 0.2\% and for a further 19\% they
differ by less than 0.35\%. Only for 40\,m baseline observations
do they differ by up to 1.4\%. These estimates justify the
assumption that $\Gamma_{1} = \Gamma_{2}$ for the purpose of
determining $\rho$ and $\theta$.  When the system is on the
meridian the angle $\psi$ in equation~(\ref{eq:cor}) is equal to
$\theta$ since the SUSI baselines are oriented North-South. In the
analysis the variation of $b$ and $\psi$ with hour angle was taken
into account as was orbital motion once a preliminary orbit had
been established for the system.  Initially $\beta$ was included
as a free parameter but after its value had been established with
additional observations as discussed in Section~\ref{sec:beta},
the fitting procedure was repeated for each night with $\beta$
fixed at the value of 0.868.

The analysis is complicated by the fact that the observed
correlation, and consequently the parameter $\gamma$, varies
during the night due to spatial seeing effects.  Depending on the
seeing conditions the data were corrected for seeing in two
different ways.

\begin{figure*}
\centerline{%
\psfig{figure=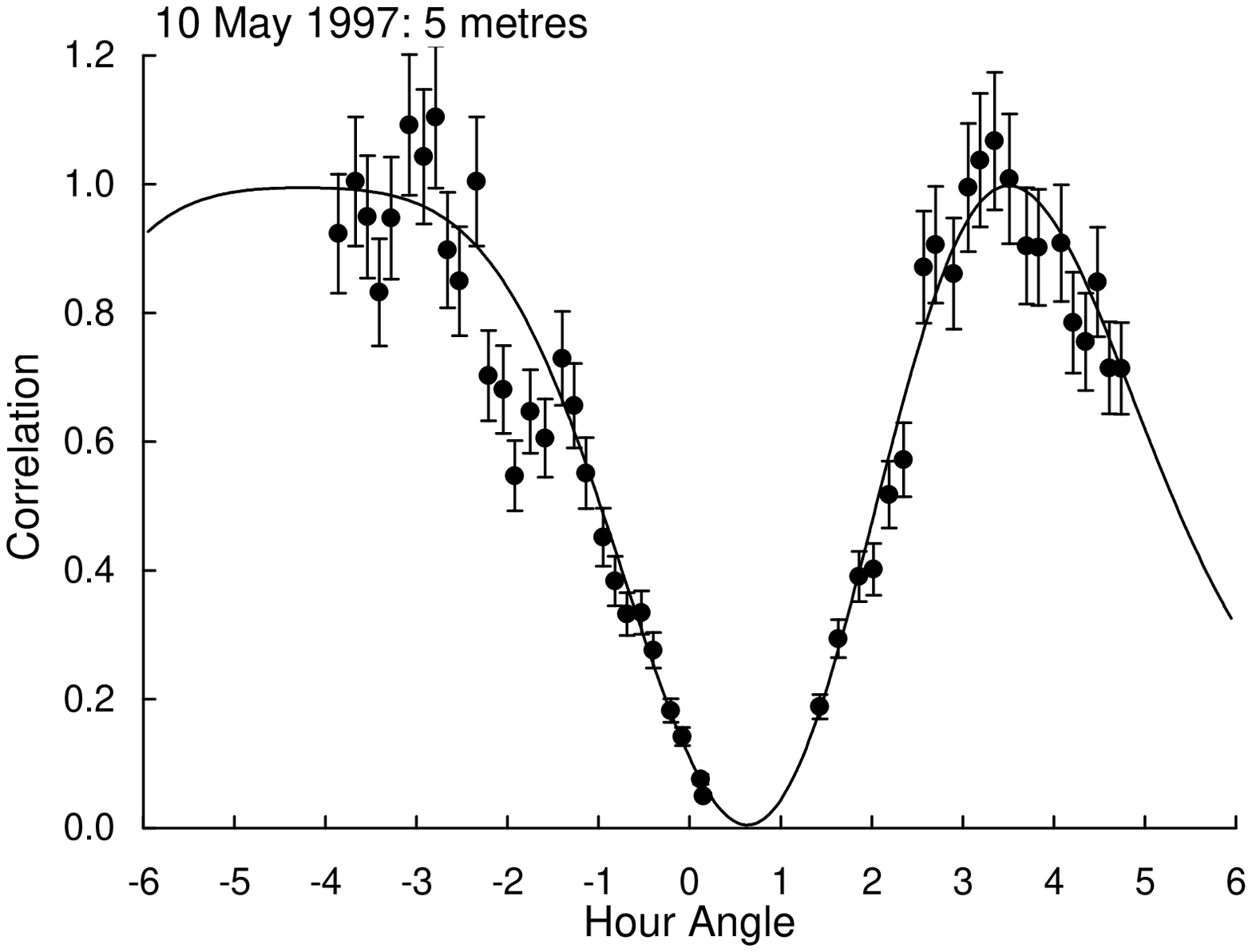,bbllx=25mm,bblly=85mm,bburx=190mm,bbury=220mm,height=60mm}
\hspace{10mm}
\psfig{figure=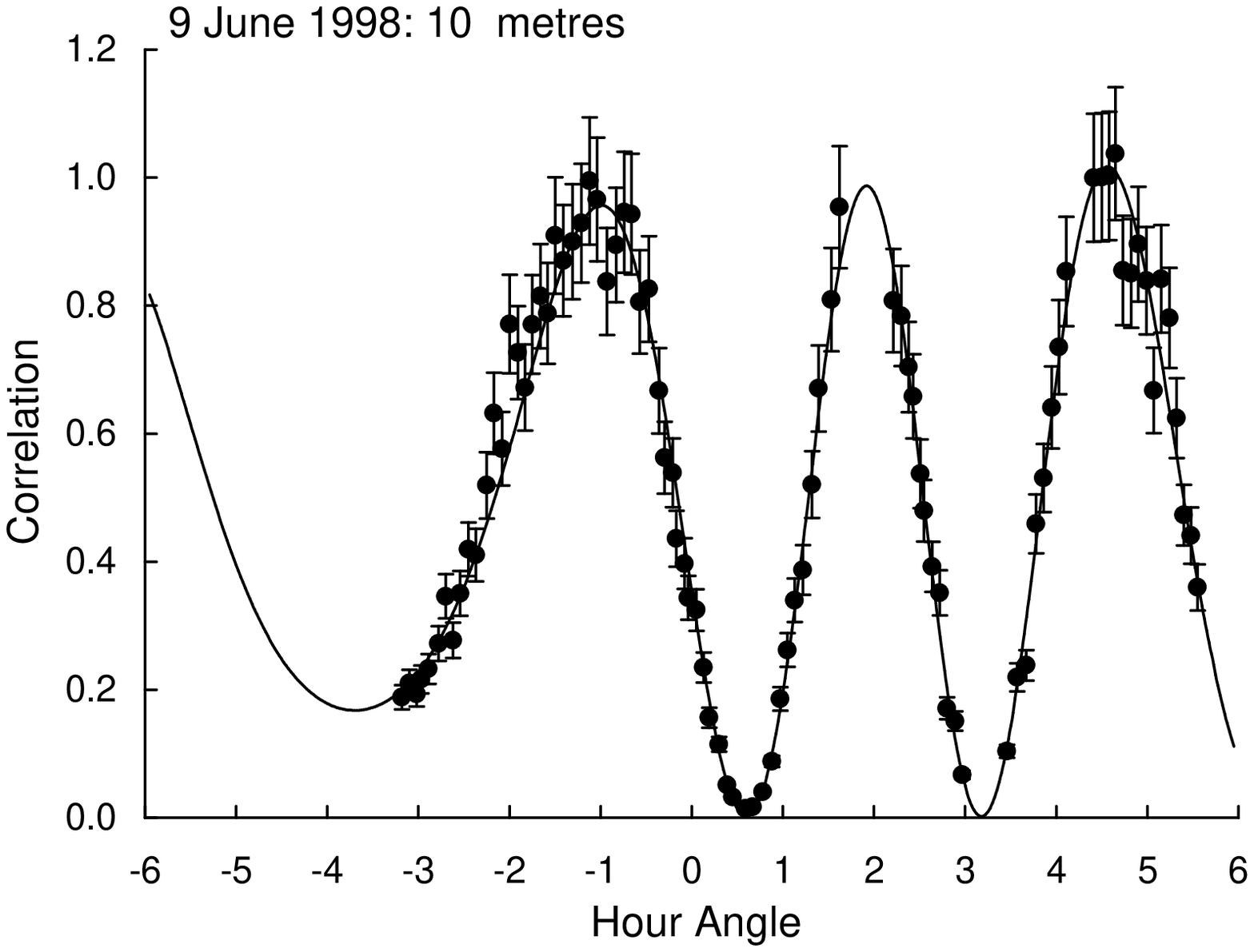,bbllx=25mm,bblly=85mm,bburx=190mm,bbury=220mm,height=60mm}}
\centerline{\hspace{5mm}(a)\hspace{81mm}(b)}
\centerline{%
\psfig{figure=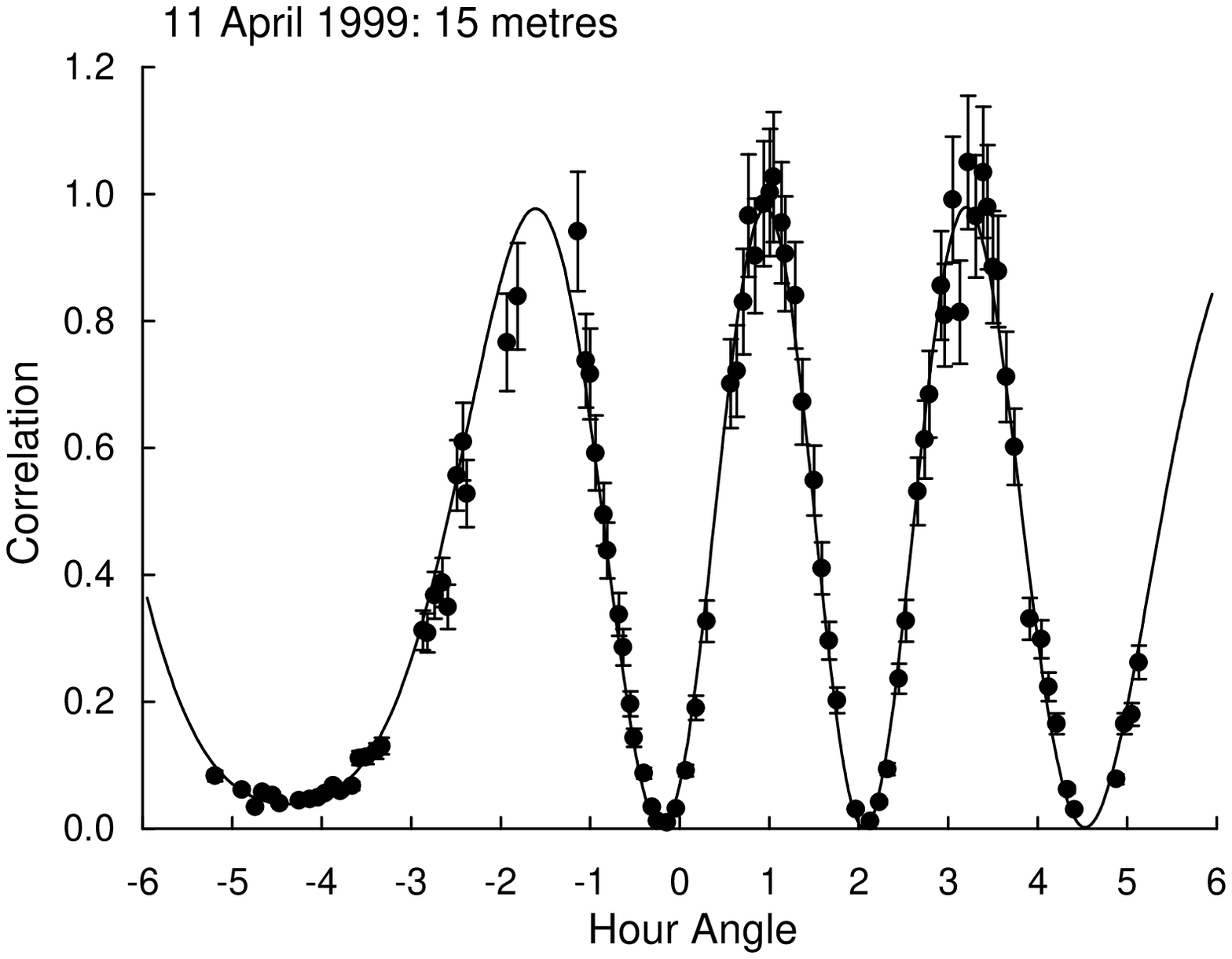,bbllx=25mm,bblly=85mm,bburx=190mm,bbury=220mm,height=60mm}
\hspace{10mm}
\psfig{figure=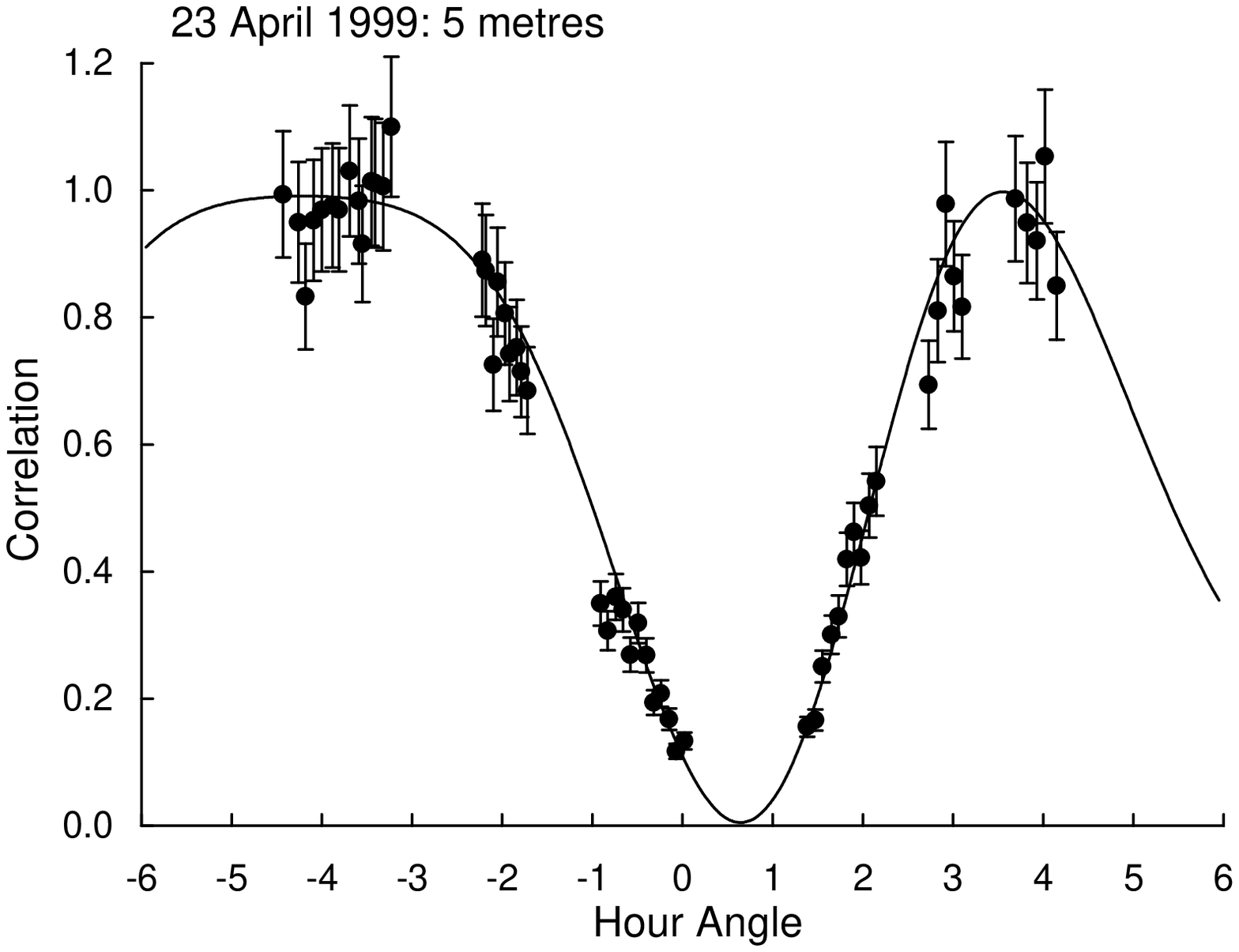,bbllx=25mm,bblly=85mm,bburx=190mm,bbury=220mm,height=60mm}}
\centerline{\hspace{5mm}(c)\hspace{81mm}(d)} \caption[Normalised
correlation for four observing nights]{Normalised correlation
versus hour angle for $\beta$ Cen on (a) May 10, 1997 at a
baseline of 5\,m, (b) June 9, 1998 at 10\,m, (c) April 11, 1999 at
15\,m, and (d) April 23, 1999 at 5\,m. In each case the filled
circles represent the measurements and the line is the fit of
equation~(\ref{eq:cor}) to the measurements.} \label{fig:dayplots}
\end{figure*}

In SUSI the spatial scale of the wavefront distortion $r_{0}$ is measured simultaneously
with the correlation \cite{99susi1}.  On some nights it was possible to
construct calibration curves of correlation versus $d/r_{0}$, where $d$ is the
aperture diameter of the interferometer, using values near maxima in the correlation
to obtain estimates of $\gamma$.  These values were corrected for the cosine term in
equation~(\ref{eq:cor}), using preliminary estimates for $\rho$ and $\theta$, and the
corrected values were plotted against $d/r_{0}$.  The data were fitted using linear
least squares (an approximate linear relationship is predicted by the standard theory
of the effects of atmospheric turbulence on an interferometer with wavefront tip-tilt
correction) and the regression line was used to calibrate all the data for the night.
The corrected data were then fitted using equation~(\ref{eq:cor}) and Fig.~\ref{fig:dayplots}
shows the results for four nights analysed in this manner in which the least squares
fitted curve is shown with the corrected data.  The error bars shown in
Fig.~\ref{fig:dayplots} are not the individually determined values for the
uncertainties in the data points since these showed large variations but were
clearly underestimates of the true uncertainty.  The error bars correspond to $\pm10\%$
in the correlation value which is believed to be the best estimate for the average
uncertainty in each point.  For the figure the maximum correlation has been arbitrarily
normalised to unity.

Inspection of Fig.~\ref{fig:dayplots} shows that the plots for 10
May 1997 and 23 April 1999, both obtained with a baseline of 5\,m,
are almost identical.  This is to be expected since they are
separated in time by 1.9973$\pm$0.0004 orbital periods (the value
adopted for the orbital period is discussed in
Section~\ref{sec:orbit}).

On some nights the correction procedure described above could not
be used, generally because there was an inadequate number of
suitable data points to construct a calibration curve.  In many of
these cases the data showed a decrease in the measured correlation
with time during the night, at least in part due to the increase
in seeing effects with hour angle.  In order to fit
equation~(\ref{eq:cor}) to these nights a linear decrease in
correlation with time was included as a free parameter in the fit.
In all cases the linear decrease in the raw correlation with time
was small and $<$0.014\,hr$^{-1}$.  A more sophisticated model for
the instrumental and seeing related trends in the correlation is
not justified for the present data. While the inclusion of the
linear decrease in the fit significantly improved the
representation of the data, it did have a small effect on the
values obtained for $\rho$ and $\theta$.  The mean difference
between the fits, with and without the linear term, is 0.29\,mas
in $\rho$ and 0.20\degr in $\theta$.  These differences are all of
the order of, or less than, the formal uncertainties in the values
from the fits and do not affect the conclusions that we draw from
the results.

Table~\ref{tab:vectors} lists the values of $\rho$ and $\theta$
determined for each night of observation.  There is a 180\degr
ambiguity in the determination of $\theta$ since the fringe phase
can not be determined with a two aperture optical interferometer
but the values listed in Table~\ref{tab:vectors} are consistent.
The uncertainties are one standard deviation values determined
using the bootstrap resampling method (see for example Babu \&
Feigelson \shortcite{96bandf}).  Due to the long period of the
orbit (of the order of a year) neither $\rho$ nor $\theta$ change
significantly during most of the nights on which observations were
made. However, an iterative procedure was adopted. A preliminary
orbital determination was used to provide the mean rates of change
of $\rho$ and $\theta$ for each night and these were then included
in the fitting procedure. In all cases the rates of change of
$\rho$ and $\theta$ did not change significantly during the course
of the night.  The mean rates of change were therefore used to
obtain a single set of values for $\rho$ and $\theta$, as listed
in Table~\ref{tab:vectors}, for the mean time of observation for
each night.

The orbital fitting program, to be discussed in
Section~\ref{sec:orbit}, accepts the position angles and
separations and allows for relative weights to be assigned to each
observational data pair.  Inspection of the plots of correlation
versus hour angle for individual nights reveals that the formal
errors listed in Table~\ref{tab:vectors} are, in many cases, not a
good guide to the weight of an individual vector pair.  Some
nights with few observational points have small formal errors
whereas other nights, with a larger number of observations, have
large formal errors.  Weights of 2, 3, 4 or 5 were therefore
assigned to each night based, not only on the quality of fit of
equation~(\ref{eq:cor}), but also taking into account the quality
and number of measured values of correlation.  The assigned
weights are listed in Table~\ref{tab:vectors}.

\subsection{The Visual Companion}\label{sec:companion}

The effect of the presence of the fainter visual companion on the
observations has been considered.  As far as is known this third
component has shown no significant motion relative to the primary
pair.  It will modulate the observed correlation, as do the two
bright components of the spectroscopic binary, but the modulation
will be small. This is partly because of the magnitude difference
but also because of the angular separation of the companion from
$\beta$~Cen.  The angular separation moves the matched path
condition for the companion away from that for $\beta$~Cen.  Thus,
for the $\beta$~Cen matched path position the correlation for the
companion is reduced by an amount that increases with baseline. An
investigation based on the Hipparcos values for the separation and
position angle of 0\farcs9 and 234\degr for the faint companion
\cite{97esa} has shown that the modulation will only be
significant at the shortest baseline of 5\,m and even there it
will not significantly affect the determination of the separation
and position angle of $\beta$~Cen.

\subsection{The Brightness Ratio} \label{sec:beta}

The ratio of the minimum correlation $C_{min}$ to the maximum
correlation $C_{max}$ at a given baseline is given by

\begin{equation}
\frac{C_{min}}{C_{max}} =
\left(\frac{\Gamma_{1}-\beta\Gamma_{2}}{\Gamma_{1}+\beta\Gamma_{2}}\right)^{2}
\label{eq:beta1}
\end{equation}

Therefore, by measuring the minimum to maximum correlation ratio,
the brightness ratio $\beta$ can be determined if $\Gamma_{1}$ and
$\Gamma_{2}$ are known.  However, if $\Gamma_{1}$ can be assumed
equal to $\Gamma_{2}$, equation~(\ref{eq:beta1}) simplifies to

\begin{equation}
\frac{C_{min}}{C_{max}} = \left(\frac{1-\beta}{1+\beta}\right)^{2}
\label{eq:beta2}
\end{equation}

Based on the estimates of the angular diameters of the component
stars, the corresponding ratio of $\Gamma_{1}$ to $\Gamma_{2}$ at
the longest baseline (15\,m) used in the determination of
$C_{min}/C_{max}$, is $>$0.998. Equation~(\ref{eq:beta2}) can
therefore be used to determine $\beta$. Observations at a baseline
of 5\,m were not included in the determination of $\beta$ because
of the potential effects of modulation due to the fainter
companion on the determination of $C_{min}$.  At longer baselines
the modulation is greatly reduced, as discussed in the previous
section, and observations are not significantly affected by it.

It was not possible to obtain a reliable value for
$C_{min}/C_{max}$ using the observations obtained in the manner
described in Section~\ref{sec:obs}.  This was because, for very
small values of correlation, the on-line correction for the
temporal effects of seeing fails. A feature of the SUSI data
handling system is that it provides measurements of the
correlation for a 1\,ms sample time ($C_{1}$), as well as the
correlation corrected for temporal seeing effects ($C$).  For
small values of correlation $C_{1}$ is more reliable than $C$.
Since $\beta$ depends critically on the correlation ratio, it was
decided to base the determination of $C_{min}/C_{max}$ on
measurements of $C_{1}$ and a modified observing approach was
adopted.   On three nights, during normal observations of
$\beta$~Cen, as minima in correlation were approached, the OPLC
was set to track the established matched path condition through
the minima and correlation measurements were taken continuously
instead of measuring delay curves.  A quadratic fit to the
observed values of $C_{1}$ through each minimum was made to
determine the minimum value of $C_{1}$.  To obtain the maximum
value of $C_{1}$, the maximum value of $C$ given by the fit of
equation~(\ref{eq:cor}) for each night was scaled by the mean
ratio of $C_{1}/C$ determined from measurements of correlation
away from the minima. The ratio $C_{min}/C_{max}$ was then
obtained from the ratio of the values of $C_{1}$ for the mimima
and maxima.  The weighted mean value for the ratio, based on the
results for 5 minima observed over three nights at baselines of
10\,m and 15\,m in April 1999, is $0.0050\pm0.0012$. Substitution
in equation~(\ref{eq:beta2}) leads to a value for $\beta =
0.868\pm0.015$.  This corresponds to a magnitude difference
between the components of $0.15\pm0.02$ at the observing
wavelength of 442\,nm. This is in agreement with the original
interferometric discovery that $\beta$ Cen is a binary with
components of approximately equal brightness \cite{74hbdanda} and
is consistent with the limit on the magnitude difference of $<0.3$
at 486\,nm reported by Robertson et al. \shortcite{99dlsb} .

\begin{figure}
\psfig{figure=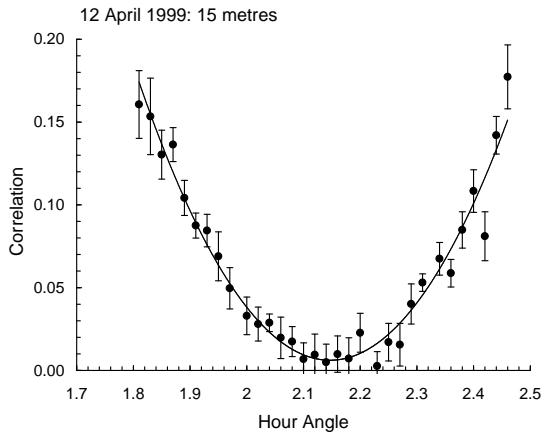,bbllx=20mm,bblly=85mm,bburx=190mm,bbury=220mm,height=60mm}
\caption{Normalised correlation at a baseline of 15\,m  versus
hour angle on April 12, 1999 through a minimum in the correlation.
The filled circles represent the measurements and the line is a
quadratic fit to the measurements.  Further details are given in
the text.} \label{fig:minimum}
\end{figure}

Fig.~\ref{fig:minimum} shows the values of correlation ($C_{1}$)
observed through a minimum on 12 April, 1999 at a baseline of
15\,m.  The correlation measures have been normalised by the value
of $C_{1}$ for the maxima on the night as explained above.

\section{Analysis of the Spectroscopic Data}

The common parameters determined in independent orbital fits to
the interferometric and spectroscopic data are the orbital period
$P$, epoch of periastron passage $T$, eccentricity $e$ and the
longitude of periastron $\omega$.  An initial orbital fit to the
interferometric data revealed small but significant differences in
the eccentricity and longitude of periastron from the
spectroscopic values published by Ausseloos  et
al.~\shortcite{aus02}.  As the value of the eccentricity is of
major importance in the derivation of accurate masses and given
that the error estimates resulting from the VCURVE code used by
Ausseloos et al. \shortcite{aus02} turn out to be too optimistic
(Harmanec, Uytterhoeven \& Aerts 2004, Uytterhoeven et al. 2004)
an attempt was made to resolve the small discrepancies by
redetermining the orbital parameters from the spectroscopic data
by using a different code.

Ausseloos et al. \shortcite{aus02} derived estimates for the
orbital elements of the primary component relative to the mass
centre by applying the L\'ehmann-Filh\'es method
\shortcite{1894lf} on the averaged radial velocities which were
determined from two Si\,III triplet lines at 4553\AA\ and 4568\AA.
Here we apply the publicly available code {\sc fotel} \cite{90had}
on the same dataset. The results are listed in the upper part of
Table~\ref{tab:orbfotel} and are to be compared with those given
in Table~3 of Ausseloos et al. \shortcite{aus02}.  As we will see
in Section~\ref{sec:orbit}, the new spectroscopic value for the
eccentricity ($e=0.819\pm0.002$) is in agreement with the
interferometric value.  The error estimates by {\sc fotel}, which
are based on a covariance matrix and should therefore be very
reliable, are slightly different from the ones stated by Ausseloos
et al. \shortcite{aus02}.

\begin{table}
\centering \caption{The orbital parameters of $\beta$\,Cen derived
  from spectroscopy with {\sc fotel} (Hadrava 1990).}\label{tab:orbfotel}
\begin{tabular}{ll} \hline
$P_{\rm orb}$ (days)       & 357.00 $\pm$ 0.07     \\
$v_{\gamma}$ (km s$^{-1}$) & 10.4 $\pm$ 3.1       \\
$K_1$ (km s$^{-1}$)        & 63.8 $\pm$ 0.6       \\
$E_0$ (HJD)                & 2451600.39 $\pm$ 0.15   \\
$e$                        & 0.819 $\pm$ 0.002     \\
$\omega$ (degrees)         & 63.6 $\pm$ 3.0        \\ \hline
$K_2$ (km s$^{-1}$)        & 63.8 $\pm$ 0.8       \\
$M_1/M_2$                  & 1.00 $\pm$ 0.04       \\  \hline
\end{tabular}
\end{table}

In order to derive the most accurate value for the mass ratio, we
also redetermined the orbital elements using the radial velocities
of both the primary and secondary, again by applying {\sc fotel}
on the same dataset as used by Ausseloos et al. \shortcite{aus02}.
These results, which are listed in the lower part of
Table~\ref{tab:orbfotel}, only differ by less than the
uncertainties in the values of the parameters from the ones
obtained previously with the L\'ehmann-Filh\'es method. A
remarkable result is the equal amplitudes of both orbits, which
implies equal masses.

\section{The Orbital Parameters} \label{sec:orbit}

The orbital parameters for the $\beta$~Cen system were initially
determined from the vector separations listed in
Table~\ref{tab:vectors} using a modification of the program
developed at the Center for High Angular Resolution Astronomy
(CHARA) \cite{89hmf}. The modified program uses the simplex
algorithm \cite{nandm65} instead of a grid search to determine the
best fitting orbit and determines the uncertainties in the orbital
elements using Monte Carlo simulation.  Both programs accept the
position angles and separations and allow for relative weights to
be assigned to each observational data pair. The programs include
corrections for precession and proper motion.  As discussed in
Section~\ref{sec:analysis}, weights of 2, 3, 4 or 5 were assigned
to each night of observation based on the quality of the fit of
equation~(\ref{eq:cor}) but also taking into account the quality
and number of measured values of correlation. The assigned weights
are listed in Table~\ref{tab:vectors}. Both programs were run with
the mean MJDs in Table~\ref{tab:vectors} converted to Besselian
epochs. There are no significant differences between the orbital
elements determined by the two methods.

\begin{table*}
\begin{minipage}{140mm}
\centering
  \caption[Orbital Parameters]{Orbital parameters for $\beta$ Cen. The first interferometric
  column is for a fit for all seven parameters; the second interferometric column is
  for a fit with the orbital period fixed at the spectroscopic value; the
  spectroscopic values are from \cite{aus02}.  Further details are given in the text.}
  \label{tab:orbpars}
  \begin{tabular}{llll}
\hline
\multicolumn{1}{c}{Parameter} & \multicolumn{1}{c}{Value} & \multicolumn{1}{c}{Value} & \multicolumn{1}{c}{Value} \\
& \multicolumn{1}{c}{(Interferometric)} & \multicolumn{1}{c}{(Spectroscopic)} & \multicolumn{1}{c}{(Interferometric)} \\
&   &    & \multicolumn{1}{c}{(Spectroscopic $P$)}  \\
\hline
Period ($P$)                          &  357\fd40$\pm$0\fd20           & 357\fd00$\pm$0\fd07  & 357\fd00   \\
Angular semi-major axis ($a$)         &  0\farcs02509$\pm$0\farcs00025 parameters&    & 0\farcs02532$\pm$0\farcs00023   \\
Inclination ($i$)                     &  67\fdg1$\pm$0\fdg5            &    & 67\fdg4$\pm$0\fdg3 \\
Epoch of periastron passage ($T$)     &  2451599\fd9$\pm$0\fd3\,(HJD)  & 2451600\fd39$\pm$0\fd15\,(HJD)  & 2451600\fd0$\pm$0\fd3\,(HJD)   \\
Position angle of line of nodes ($\Omega$) & 288\fdg5$\pm$0\fdg4       &    & 288\fdg3$\pm$0\fdg4 \\
Eccentricity ($e$)                      &  0.821$\pm$0.005             & 0.819$\pm$0.002      & 0.824$\pm$0.004    \\
Longitude of periastron ($\omega$)    &  60\fdg8$\pm$0\fdg6            & 63.6$\pm$3.0         & 61\fdg3$\pm$0\fdg6  \\
\hline
\end{tabular}
\end{minipage}
\end{table*}

The orbital parameters obtained from the fit to the 43 entries in
Table~\ref{tab:vectors}, using the modified program, are listed in
Table~\ref{tab:orbpars}. The common parameters determined
spectroscopically and listed in Table~\ref{tab:orbfotel} are
repeated in Table~\ref{tab:orbpars}.  The common parameters are in
good agreement but we note that the orbital period from the
spectroscopic solution has a much smaller uncertainty than that
determined interferometrically.  This is to be expected since the
spectroscopic observations spanned $\sim$ 12 years compared with
$\sim$ 5 years for the interferometric observations.  Following
the example set by Herbison-Evans et al. \shortcite{spica} in
their analysis of interferometric observations of $\alpha$~Vir,
the orbital period was fixed at the spectroscopic value and the
orbit fitting program was run again on the interferometric data.
The resulting values for the remaining six parameters are listed
in Table~\ref{tab:orbpars}.  The differences between the
parameters, with and without the period fixed, are generally less
than, or of the order of the uncertainties in the values of the
parameters. The fit with the period fixed is marginally better
than with the period a free parameter and, although it is not
evident in the rounded values shown in Table~\ref{tab:orbpars},
the uncertainties in the derived parameters were all smaller with
the period fixed. For this reason and the longer span  of the
spectroscopic observations, the orbital parameters for the fit
with the orbital period fixed are adopted as best representing the
interferometric data.

\begin{figure}
\resizebox{\hsize}{!}{\includegraphics{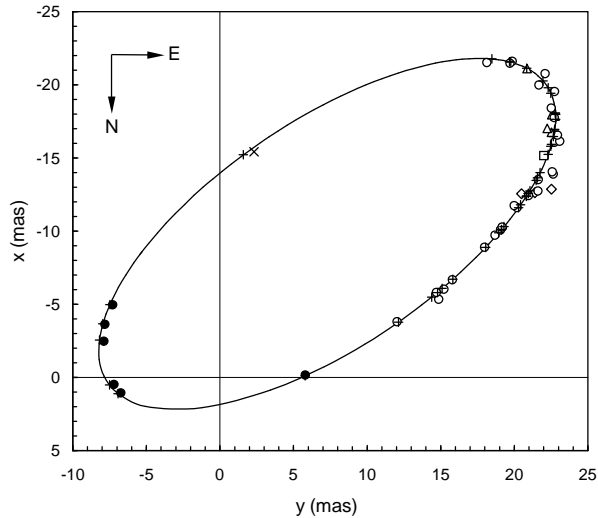}}
  \caption{The orbit of $\beta$ Centauri. The data
points are the SUSI results from Table~\ref{tab:vectors} except
for the MAPPIT point which is from Robertson et al. (1998). SUSI
observations: 1997 $\diamondsuit$; 1998 $\triangle$; 1999 {\Large
$\circ$}; 2000 {\Large $\bullet$}; 2002 {\bf $\sq$}. MAPPIT
observation: 1995 {\large\textsf{x}}. The points on the fitted
orbit that correspond to the observational points are marked with
$\large +$.  Further details are given in the text.}
  \label{fig:orbit}
\end{figure}

The observationally determined orbital points and the orbit for
the adopted orbital parameters are plotted in Fig:~\ref{fig:orbit}
following the common double-star convention \cite{78wdh}.  Also
plotted are the positions on the fitted orbit for the epochs of
the individual observational points.  Error ellipses for the
observational points have not been plotted since they are
comparable in size or smaller than the symbols representing the
observational points and, as discussed in
Section~\ref{sec:analysis}, they are not always the best
representation of the weight of an individual point. We note that
the orbit with the orbital period as a free parameter is almost
indistinguishable from the plotted orbit, except that many of the
observational points lie a little further from the corresponding
positions on the orbit. This is further evidence supporting the
decision to accept the orbital fit with the period fixed.

In Fig.~\ref{fig:orbit} the vector position determined with MAPPIT
by Robertson et al. \shortcite{99dlsb} has been plotted and, while
it was not included in the fit, it is seen to be consistent with
the orbit derived from the SUSI data. There is a 180\degr
ambiguity in the position angle of the MAPPIT point relative to
the SUSI data and in the figure the point has been plotted at
position angle 172\degr rather than the value of 352\degr given by
Robertson et al.

\begin{figure}
\resizebox{\hsize}{!}{\includegraphics{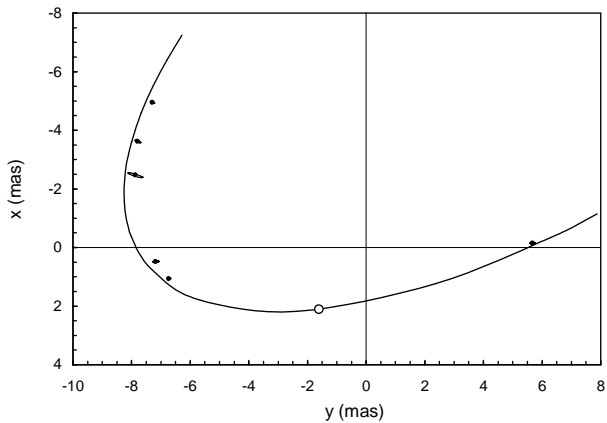}}
  \caption{A section of the orbit of $\beta$ Centauri around periastron. The data
points are the SUSI results from Table~\ref{tab:vectors}. Each
point has its error ellipse plotted but, in all but one case, it
is too small to extend outside the plotted point.  The curve is
the orbital fit shown in Fig.~\ref{fig:orbit}.  The location of
periastron is marked by a circle.}
  \label{fig:peri}
\end{figure}

Examination of Fig.~\ref{fig:orbit} reveals that, while over most
of the orbit the observational points are distributed on either
side of the curve, the observations made in 2000 near to
periastron all lie inside the fitted orbit.  This is shown on an
expanded scale in Fig.~\ref{fig:peri} where a section of the orbit
close to periastron is plotted on an enlarged scale.  While we
have not been able to provide an explanation for this phenomenon
we conjecture that, since the separation of the component stars is
less than 7 stellar radii at periastron, the possibility of
interaction may affect the brightness distributions such that the
components appear closer than their centres of mass.
Miroshnichenko et al. \shortcite{01metal} have reported
interaction around periastron for the early B type binary $\delta$
Sco involving a Be outburst. Like $\beta$ Cen, $\delta$ Sco has a
highly eccentric orbit, but with an orbital period of $\sim$ 10.6
years. Although $\delta$ Sco spends longer close to periastron,
the closest approach is of the order of 30 times the primary
radius compared with less than 7 radii for $\beta$ Cen. This
suggests that some form of interaction is likely to occur.

\section{Combination of the Spectroscopic and Interferometric
Results}\label{sec:results}

The interferometric and spectroscopic results can be combined to
yield the distance to the system, the absolute magnitudes of the
components and the masses of the components.  Not all the
parameters are required for this purpose and the relevant
parameters from the two techniques are summarised in
Table~\ref{tab:params}. A mean of the interferometric and
spectroscopic values for the eccentricity is adopted.

\begin{table}
\centering
  \caption[$\beta$ Cen parameters]{Values of parameters for $\beta$ Cen
  determined directly by interferometry and spectroscopy and used to derive the
  basic physical parameters of the system. }
  \label{tab:params}
  \begin{tabular}{lll}
\hline \multicolumn{1}{c}{Parameter} & \multicolumn{1}{c}{Value} &
\multicolumn{1}{c}{Source}\\
\hline Period ($P$)   & 357\fd00$\pm$0\fd07   & S \\
$K_{1}$ (km s$^{-1}$)       & 63.8$\pm$0.6      & S \\
$K_{2}$ (km s$^{-1}$)       & 63.8$\pm$0.8      & S \\
Eccentricity ($e$)            & 0.821$\pm$0.003   & I + S\\
Angular semi-major axis ($a$) & 0\farcs02532$\pm$0\farcs00023 & I \\
Inclination ($i$)   & 67\fdg4$\pm$0\fdg3      & I \\
Brightness ratio ($\beta$) (at $\lambda$442\,nm) & 0.868$\pm$0.015  & I \\
\hline
\end{tabular}
\end{table}

As the two sets of data result in a single self-consistent set of
parameters we are able to derive the masses of the components with
high precision.  Functions containing the masses of the primary
(M$_{1}$) and secondary (M$_{2}$) components in solar mass units
and the semi-major axis ($a$) of the orbit in AU are given by
\cite{78wdh}:

\begin{equation}
M_{1}\sin^3{i} =
1.036\times10^{-7}(K_{1}+K_{2})^{2}K_{2}P(1-e^{2})^{3/2},
\label{eq:m1}
\end{equation}

\begin{equation}
M_{2}\sin^3{i} =
1.036\times10^{-7}(K_{1}+K_{2})^{2}K_{1}P(1-e^{2})^{3/2},
\label{eq:m2}
\end{equation}

\begin{equation}
a\sin{i} = 9.1919\times10^{-5}(K_{1}+K_{2})P(1-e^{2})^{1/2}.
\label{eq:sma}
\end{equation}

The values for these functions, computed from the data in
Table~\ref{tab:params}, are listed in Table~\ref{tab:derived}.

\begin{table}
\centering
  \caption[Derived parameters for $\beta$ Cen]{Values of parameters for $\beta$
  Cen derived from the parameters listed in Table~\ref{tab:params} using
  equations (1)--(3).  $M_{1}$ and $M_{2}$ are the masses of the primary and
  secondary respectively and $a$ is the semi-major axis of the orbit.  }
  \label{tab:derived}
  \begin{tabular}{lcl}
\hline \multicolumn{1}{c}{Parameter} & \multicolumn{1}{c}{Value} &
\multicolumn{1}{c}{Reference} \\
\hline
$M_{1}\sin^{3}{i}$ (M$_\odot$) & 7.15$\pm$0.24 & Equ.~(\ref{eq:m1})  \\
$M_{2}\sin^{3}{i}$ (M$_\odot$) & 7.15$\pm$0.23 & Equ.~(\ref{eq:m2})  \\
$a\sin{i}$ (AU)   & 2.391$\pm$0.032 & Equ.~(\ref{eq:sma})  \\
\hline
\end{tabular}
\end{table}

\subsection{The distance to $\beta$ Cen}\label{sec:parallax}

The parallax of the system can also be determined by combining the
interferometric and spectroscopic results and is given by:

\begin{equation}
\pi\arcsec = \frac{a\arcsec\sin{i}}{(a\sin{i})} \label{eq:dist}
\end{equation}

A parallax determined in this way has been appropriately termed an
orbital parallax \cite{92orbpar}. Combining $a\arcsec$ and $i$
from Table~\ref{tab:params} with $a\sin{i}$ from
Table~\ref{tab:derived} gives the orbital parallax of $\beta$~Cen
equal to 9.78$\pm$0.16\,mas.  This is to be compared with the
HIPPARCOS parallax of 6.21$\pm$0.56\,mas \cite{97esa}.  There is a
large and significant difference in spite of the relatively large
fractional uncertainty in the HIPPARCOS value.  The HIPPARCOS
catalogue recognises $\beta$ Cen as a binary system but has no
mention of the dual nature of the primary component of $\beta$ Cen
and lists only the faint third magnitude companion.  We conjecture
that the HIPPARCOS value for the parallax is seriously affected by
the binary nature of the primary of $\beta$ Cen. In fact, the
HIPPARCOS observations of spectroscopic binaries were not fitted
in general with an orbital model to derive the value for the
parallax.  While reprocessing of the HIPPARCOS astrometric data
has been performed for many single-lined spectroscopic binaries
(see for example \cite{pandb03} and references therein), it has
not been done for $\beta$ Cen.  For this reason we adopt the value
derived directly from the interferometric and spectroscopic data.
The parallax we have adopted is equivalent to a distance of
102.3$\pm$1.7\,pc.

\subsection{The absolute magnitudes and spectral types of the
components of $\beta$ Cen}\label{sec:sp}

Breger \shortcite{67breger} has given the following photometric
data for $\beta$~Cen: \textit{V} = 0.61, (\textit{B-V}) = -0.24
and (\textit{U-B}) = -0.99. Since the magnitude difference is only
0.15$\pm$0.02 at a wavelength of 442\,nm (from the brightness
ratio of 0.868$\pm$0.015 given in Table ~\ref{tab:params}) and, as
will be shown in Section~\ref{sec:masses}, the masses of the
components are equal, any difference in the colours of the two
components will be very small and they will be assumed to be the
same. Reddening will be small for this relatively nearby system
(distance $\sim$ 102\,pc) and \textit{E(B-V)} is estimated to be
$\sim$ 0.03. It follows that (\textit{B-V})$_{0}$ = -0.27 and we
adopt \textit{A}$_{V} = 0.09\pm0.03$.

Using the parallax from Section~\ref{sec:parallax} the absolute
visual magnitude $M_{V}$ of the system is given by:

\begin{equation}
M_{V} = V - 0.09 - 5.0\log (102.4/10). \label{eq:absmag}
\end{equation}

Thus $M_{V}$ = -4.53$\pm$0.05 for the two components combined. The
brightness ratio of 0.868$\pm$0.015, although measured at
$\lambda$442\,nm, is unlikely to differ significantly in the
\textit{V} band because of the equality of their masses and the
small difference in luminosity between the two components.
Assuming that the same brightness ratio is valid in the visual
band, the absolute visual magnitudes for the two components are
$M_{V\mathrm{primary}} = -3.85\pm0.05$ and
$M_{V\mathrm{secondary}} = -3.70\pm0.05$.

These values for the absolute visual magnitudes lie between the
values given by Balona \& Crampton \shortcite{74bandc} for
spectral types B1\,III (-4.1) and B2\,III (-3.7) and are close to
the value of -3.9 given by Schmidt-Kaler \shortcite{landb} for
B2\,III.  They lie in the middle of the range of absolute visual
magnitudes for $\beta$\,Cep stars \cite{93sandj}.

\subsection{The masses of the components of
$\beta$~Cen}\label{sec:masses}

The combination of the value for $i$ from Table~\ref{tab:params}
with the values for $M_{1}\sin^3{i}$ and $M_{2}\sin^3{i}$ from
Table~\ref{tab:derived} gives the mass of the primary $M_{1}$
equal to 9.09$\pm$0.32\,M$_\odot$.  Similarly the mass of the
secondary $M_{2}$ is 9.09$\pm$0.31\,M$_\odot$. These are the most
accurate mass estimates of any $\beta\,$Cep star in a binary
system known to date.

Masses for early type giants are in general rather uncertain.  The
mass for a B0\,III type star is usually taken as $\sim$
20\,M$_\odot$ and, for a B5\,III type, $\sim$ 7\,M$_\odot$
\cite{landb}.  The values we have determined for the components of
$\beta$ Cen are significantly lower at B1\,III than implied by
those generally accepted for B giant stars.

Refined mass estimates at $\sim$3.5 per cent accuracy, which is
the level of our results for $\beta\,$Cen, have recently become
available for two single $\beta\,$Cep stars from asteroseismology,
i.e.\ from detailed interpretation of the measured frequencies and
mode identification of their oscillations. Such studies have led
to very similar mass estimates for these two single stars, i.e.\ a
mass range [9.0,9.5]\,M$_\odot$ for HD\,129929 \cite{03aerts} and
[9.0,9.9]\,M$_\odot$ for $\nu\,$Eri \cite{04phd}. The mass of the
eclipsing binary $\beta\,$Cep star 16\,Lac is also rather well
constrained and lies in [9.0,9.7]\,M$_\odot$ \cite{03thoul}.

The masses of all the known $\beta\,$Cep stars as a group of
variables have mainly been derived from photometric calibrations,
e.g.\ of the Walraven, Geneva and Str\"omgren systems.  This
method provided systematically too high masses for the three stars
with accurate seismic mass determination: $M\simeq\,10$ to
11.3\,M$_\odot$ for HD\,129929; $M\simeq\,10$ to 10.8\,M$_\odot$
for $\nu\,$Eri and $M=10\,$M$_\odot$ for 16\,Lac (values taken
from Heynderickx, Waelkens \& Smeyers 1994; see also Sterken \&
Jerzykiewicz 1993) to be compared with the seismic values listed
above.  The situation is of course worse when we consider the mass
estimates of photometric calibrations for spectroscopic binaries
since the binarity is not usually taken into account in such
global calibrations.  Heynderickx et al. \shortcite{94hws}, for
example, list a mass between 14.4 and 15.2\,M$_\odot$ from
treating $\beta\,$Cen as a single star, as they did all other
stars in their sample.  Their estimate for the mass is far too
high compared with our results.  Since a large fraction of the
$\beta\,$Cep stars turn out to be a member of a spectroscopic
binary with not too different components (Aerts \& De Cat 2003),
it is unavoidable that the mass estimates of these group members
from photometry are very unreliable.

\begin{table}
\centering
  \caption[Results for $\beta$ Cen]
{The parallax ($\pi$), masses ($M_{1}$ \& $M_{2}$) and
  absolute visual magnitudes ($M_{1 V}$ \& $M_{2 V}$) for $\beta$ Cen and
  its component stars.  The absolute
  visual magnitudes were computed directly from the photometric data,
  the adopted parallax, and the brightness ratio $\beta$ given in
  Table~\ref{tab:params}. }
  \label{tab:results}
  \begin{tabular}{lr}
\hline \multicolumn{1}{c}{Parameter} & \multicolumn{1}{c}{Value} \\
\hline
$\pi$ (mas)                 & 9.77$\pm$0.15  \\
$M_{1}$ (M$_\odot$)        & 9.09$\pm$0.31  \\
$M_{2}$ (M$_\odot$)        & 9.09$\pm$0.30  \\
$M_{1 V}$                   &  -3.85$\pm$0.05 \\
$M_{2 V}$                   &  -3.70$\pm$0.05\\
\hline
\end{tabular}
\end{table}

\section{Summary}\label{sec:sum}

The orbital parameters for the primary component of $\beta$ Cen
have been determined from interferometric observations made with
SUSI.  These parameters have been combined with the high
resolution double-lined spectroscopic data for $\beta$~Cen
reported in Ausseloos et al. \shortcite{aus02}, the analysis of
which has been revised here. The combination of data has yielded
the masses and absolute visual magnitudes of the component stars
and an accurate distance to the system. The very precise results
of this powerful combination of techniques are summarised in
Table~\ref{tab:results}.

The results presented in this paper constitute a very suitable
starting point for any future asteroseismic modelling of
$\beta\,$Cen's two pulsating components. The star is also an ideal
target to investigate possible tidal effects on the oscillations
in view of the very close passage (less than 10 times the radii of
the component stars) of the stars at periastron although, for
this, long-term monitoring over several orbits is required.

\section{Acknowledgements}

The SUSI programme has been funded jointly by the Australian
Research Council and the University of Sydney, with additional
support from the Pollock Memorial Fund and the Science Foundation
for Physics within the University of Sydney.  The authors are
grateful to Bill Hartkopf for making the binary orbit fitting
program available to us and for assistance with its
implementation.  MA, CA and KU thank Petr Hadrava and Petr
Harmanec for putting the FOTEL code at their disposal and
acknowledge financial support from the Research Fund K.U.Leuven
under grant GOA/2003/04.

\end{document}